\documentclass[12pt]{revtex4}
\usepackage{amssymb}
\usepackage{graphicx}

\def\be{\begin{equation}}
\def\ee{\end{equation}}

\def\be{\begin{equation}}
\def\ee{\end{equation}}

\newcommand{\lsim}{\raisebox{-.4ex}{$\stackrel{<}{\scriptstyle \sim}$}}
\newcommand{\gsim}{\raisebox{-.4ex}{$\stackrel{>}{\scriptstyle \sim}$}}

\begin{document}
\title{ Can the viscosity in astrophysical black hole accretion disks be close to its string theory bound? }
\smallskip\smallskip 
\author{Banibrata Mukhopadhyay}
\affiliation{Department of Physics, Indian Institute of
Science, Bangalore 560012, India}
\email{bm@physics.iisc.ernet.in}  

\vskip10cm 
\begin{abstract}
String theory and gauge/gravity duality suggest the lower bound of shear viscosity
($\eta$) to entropy density ($s$) for any matter to be $\sim \mu\hbar/4\pi k_B$, when
$\hbar$ and $k_B$ are reduced Planck and Boltzmann constants respectively and $\mu\le 1$. 
Motivated by this, we explore $\eta/s$ in black hole accretion flows,
in order to understand if such exotic flows could be a natural site for the
lowest $\eta/s$.
Accretion flow plays an important role in black hole physics in identifying the 
existence of the underlying black hole. This is 
a rotating shear flow with insignificant molecular viscosity, which could however have a significant turbulent 
viscosity, generating transport, heat and hence entropy in the flow. However, in presence of strong
magnetic field, magnetic stresses can help in transporting matter independent of
viscosity, via celebrated Blandford-Payne mechanism. In such cases, energy and then
entropy produces via Ohmic dissipation. In addition, certain optically thin, hot, accretion
flows, of temperature $\gsim 10^9$K, may be favourable for nuclear burning which could 
generate/absorb huge energy, much higher than that in a star.
We find that $\eta/s$ in accretion flows appears to be close to the lower bound suggested 
by theory, if they are embedded by strong magnetic field or producing nuclear energy,
when the source of energy is not viscous effects.
A lower bound on $\eta/s$ also leads to an upper bound on the Reynolds number of the flow. 
\end{abstract}

\maketitle



\newpage
\section{Introduction}\label{intro}
In order to explain the Quark Gluon Plasma (QGP) observables at Relativistic Heavy Ion Collider (RHIC) having
temperature ($T$) $\gsim 10^{12}$K, 
it has become apparent that a non-zero shear viscosity, $\eta$, is needed. One way of characterizing $\eta$, 
which is a dimensionful number, is to take its ratio with the entropy density, $s$. It turns out that  
RHIC plasma has
\be\label{ebs}
\frac{\eta}{s}\sim 0.1 \frac{\hbar}{k_B}\,.
\ee
Although $\eta$ itself is large (in CGS units $\sim 10^{12}$ gm/cm/s), its ratio with $s$ is small. 
Note that this ratio for a known fluid, say water, is about two orders of magnitude greater than the above value. 
Hydrodynamic simulations at RHIC and Large Hadron Collider (LHC) (which provide the environments created 
around a few microseconds after the big bang) as well suggest a small value for this ratio in the hot QGP state. 
Such a small value points towards strongly 
interacting matter. In fact, string theory arguments and the famous gauge/gravity duality in the form of 
the AdS/CFT correspondence suggest a lower bound \cite{kss} for $\eta/s$ for any matter given by
\be
\frac{\eta}{s}\geq \frac{\mu}{4\pi} \frac{\hbar}{k_B} \sim 0.08 \mu  \frac{\hbar}{k_B}\,,
\label{etabys}
\ee
where $\mu<1$ \cite{bms} but non-zero \cite{bek,teaney} (see, e.g., \cite{bnd}).
For the present purpose, we assume that there
is a bound and that $\mu \sim 1$. 
 

Is there any natural site revealing an $\eta/s$ close to above lower bound? 
Note that evenif $\mu<<1$, still the same question remains.
Such a value may be possible to
arise naturally, if the temperature and/or density of the systems 
are/is same as that in RHIC/LHC. In the search of such an exotic system, our story starts which could
serve as a natural verification of gauge/gravity
duality in the form of the AdS/CFT correspondence.

Observed data suggest that the accretion flows around certain black holes must be optically thin, 
geometrically thick and very hot. Such flows often exhibit outflows/jets in their hard spectral states.
The ion temperature ($T_i$) in such flows/disks is  
as high as $7\times 10^{11}$K. 
The supermassive black hole system at the centre of
our galaxy, Sgr~A$^*$ \cite{rameshnature}, (most of the temporal classes of) the X-ray binary
Cyg~X-1 \cite{sl75} are some of the examples of this kind of hot accretion disks, which  
exhibit radiatively inefficient flows. They are particularly different from their
radiatively efficient counter part, namely the Keplerian accretion disk, which 
is geometrically thin, optically thick, cooler, and of temperature $T\sim 10^7$K \cite{ss73}.
The flow in certain temporal classes (and soft spectral states) of micro-quasar GRS~1915+105 \cite{remillard} is an 
example of Keplerian disk. While the temperature of the latter cases could be 
same order in magnitude as that in the center of a star,
due to much low density they practically do not exhibit any thermonuclear burning. Hence any source of 
energy therein must be due to magnetic and viscous effects. On the other hand, the former cases may be
favourable for thermonuclear reactions due to their very high temperature, 
evenif their 
density is low \cite{mc00,mc01}. Hence, the energy released/absorbed due to nuclear
reactions in an optically thin, hot, accretion flow could be comparable to or even dominating over
its viscous counter part \cite{cm99,mc00,mc01}. Being very hot, such flows are also highly
ionized and hence expected to be strongly magnetized (see e.g. \cite{ghise}) rendering magnetic dissipation.
In fact, the celebrated Blandford-Payne mechanism \cite{bland}) is based on such a magnetized
accretion disk in the Keplerian regime, which argues for angular momentum transport due to
outflows/jets through the outgoing magnetic field lines, in absence of shear (turbulent) viscosity.
Now an obvious question arises: what is the value of $\eta/s$ in
optically thin accretion flows (e.g. Advection Dominated Accretion Flow
(ADAF) \cite{ny94}, General Advective Accretion Flow (GAAF) \cite{mukhraj10})
when the temperature is close to that of the QGP matter in RHIC? Primarily guess is that it
is not too much different from its lower bound as in equation (\ref{ebs}). Is this that simple to anticipate?
Note that accretion flows with very small molecular viscosity (and then very large Reynolds number) 
exhibit turbulence and then turbulent viscosity. 
If the turbulent viscosity dominates, then the energy dissipated due to viscosity 
in disks is comparable/dominant to/over
the magnetic/nuclear energy, as will be discussed in detail below.
Note also that at low densities (zero chemical
potential), lattice QCD calculations \cite{fodor1,qcd1,qcd2} suggest a crossover from the hadronic state
to the QGP state at around $1.5–2 \times 10^{12}$K \cite{fodor2}. At such high temperatures,
there is expected to be copious pion production in a strongly interacting
system. This QGP state is thought to have existed a few microseconds
after the big bang and is being extensively studied at RHIC and LHC. Of course
the density in early universe is large as well, unlike that in
accretion flows, a large temperature is well enough to reveal such a
phase which has been mimicked in RHIC/LHC.
However, at a very large density (several orders of magnitude larger than that in 
accretion flows) the lattice QCD results are not valid (Zoltan Fodor, private communication).

Recently, Sinha \& Mukhopadhyay \cite{sm11} initiated to look at the above issue and 
argued that the Shakura-Sunyaev turbulent viscosity parameter $\alpha$ \cite{ss73} should not be constant
throughout the flow, and be decreasing
with the increase of temperature and/or density of the flow. Based on the general relativistic
model of a viscous ADAF, they 
showed that $\eta/s$ for an astrophysical black hole,
at finite $\alpha$, is always several
orders of magnitude higher than its value in a QGP fluid. Then they argued for the flow to become
very weekly viscous (not exhibiting turbulence) close to a black hole, rendering a smaller $\eta/s$. The
results appear to be independent of the choice of equation of state (EoS); whether of ideal gas or
of QCD which is appropriate for flows with $T\gsim 10^{12}$K and low density (close to zero chemical 
potential). However, they could not clarify the natural circumstances when $\alpha$ is small.
Note that small $\alpha$ would reveal negligible dissipation of energy in the flow. Hence,
question arises what is the source of energy and entropy in the accretion flows, which is also
very important in order to address $\eta/s$? In addition, how is the accretion possible in a
very small-$\alpha$ flow? Moreover, as will be explained in the next section, above 
work did not model the entropy of the flow adequately, while generally interpreted the
results correctly. Instead, that was the first step forward in the direction of evaluating 
$\eta/s$ for any astrophysical flow, eventhough incomplete, which is a potential
natural site for small $\eta/s$. In the present work, we have
removed the above uncertainties lying in the previous work. Then we plan to establish the underlying 
physics giving rise to the astrophysical values of $\eta/s$. More precisely, we plan to
address: (1) regime of accretion flows giving rise to $\eta/s$
close to its theoretical lower bound, (2) observational implications of such regimes, 
(3) black hole sources presumably revealing lower bound of $\eta/s$, 
(4) constraining physical parameters (e.g. Reynolds number) of accretion flows based on $\eta/s$ bound.

In the next section, we discuss the salient features of hot advective 
accretion flows and the importance of strong magnetic field and 
nuclear energy produced therein in order to obtain small $\eta/s$. Subsequently in \S 3 we analyse the
solutions of the equations and address the flow parameters making $\eta/s$ to be close to
the lower limit. Finally \S 4 summarizes our findings with discussion and implication.


\section{Advective accretion flows around black holes}\label{gaaf}


In order to understand the underlying physics, let us consider the equations describing 
optically thin, viscous, magnetized, advective accretion disk \cite{mg03}. 
We describe the model in the pseudo-Newtonian framework with the Paczy\'nski-Wiita potential \cite{pw80,m02}.
This is particularly because, for the present purpose, the pure general relativistic results
for a rotating black hole qualitatively do not reveal any new physics. Hence, the vertically averaged hydromagnetic
equations of energy-momentum balance along with the equations of continuity, induction and divergence 
of magnetic field in the limit of very large conductivity are
\begin{eqnarray}
\nonumber
&&\dot{M} = −4\pi x\Sigma\vartheta,\\
\nonumber
&&\vartheta\frac{d\vartheta}{dx}+\frac{1}{\rho}\frac{dP}{dx}-\frac{\lambda^2}{x^3}+\frac{1}{2(x-1)^2}=0,\\
\nonumber
&&\vartheta\frac{d\lambda}{dx}=\frac{1}{x\Sigma}\frac{d}{dx}\left(x^2W_{x\phi}\right)+\frac{hx}{\Sigma}\left(B_x\frac{dB_\phi}{dx}+
B_z\frac{dB_\phi}{dz}+\frac{B_xB_\phi}{x}\right),\\
\nonumber
&&\vartheta hT\frac{ds}{dx}=\frac{\vartheta h}{\Gamma_3-1}\left(\frac{dP}{dx}-\frac{\Gamma_1 P}{\rho}\frac{d\rho}{dx}\right)=Q^+-Q^-,\\
\nonumber
&&\frac{d}{dx}\left(xB_x\right)=0,\\
\nonumber
&&\frac{d}{dx}\left(\vartheta B_\phi-\frac{B_x\lambda}{x}\right)=0,\\
&&\frac{d}{dx}\left(x\vartheta B_z\right)=0,
\label{eqnset}
\end{eqnarray}
assuming that the variables do not vary significantly in the vertical direction such that $d/dz\rightarrow 1/z$, 
which is indeed true for the disk flows.
Here $\dot{M}$ is the conserved mass accretion rate, $\Sigma$ and $\rho$ are the vertically integrated density and
density of the flow respectively, $\vartheta$ is the radial velocity, $P$ the total pressure, $\lambda$ the
specific angular momentum, $W_{x\phi}$ the vertically integrated shearing stress, $h$ the half-thickness, 
$s$ the entropy per unit volume, $T$ the temperature of the flow, $Q^+$ and $Q^-$ 
are the vertically integrated net energy released and absorbed rates in/from the flow
respectively, $\Gamma_1, \Gamma_3$
indicate the polytropic indices depending on the gas and radiation content in the flow (see, e.g., \cite{mukhraj10} 
for exact expressions) and $B_x$, $B_\phi$ and $B_z$ are 
the components of magnetic field. 
Note that, the independent
variable $x$ is the radial coordinate of the flow expressed in the units of $2GM/c^2$, where $G$ is the 
gravitation constant, $M$ the mass of the black hole and $c$ the speed of light. 
Accordingly, all the above variables are expressed in dimensionless units. For any other
details, see the existing literature \cite{mukhraj10-2,mukhraj10}.

Let us now pay a special attention to the energy equation of the equation set (\ref{eqnset}). We can rewrite it as
\be
\vartheta T\frac{ds}{dx}=q_{\rm vis}+q_{\rm mag}+q_{\rm Nex}-q_{\rm rad}-q_{\rm Nend},
\label{modion}
\ee
where $q_{\rm vis}$, $q_{\rm mag}$ and $q_{\rm Nex}$ respectively define the energies released per unit volume 
per unit time due to
viscous dissipation, magnetic dissipation and thermonuclear reactions and $q_{\rm rad}$ and $q_{\rm Nend}$ respectively indicate  
the energy radiated away per unit volume per unit time by various cooling processes
like Bremsstrahlung, synchrotron and inverse-Comptonization of soft photons, and the 
energy absorbed per unit volume per unit time due to thermonuclear reactions. 

It was shown earlier that the nuclear energy released/absorbed in a hot accretion disk need
not be negligible \cite{cm99,mc00,mc01,zhang09} and the thermonuclear reactions therein could produce 
over-abundant metals \cite{mc00,tao08,tao11,filho03} which are observed in galaxies. However, the value
of $\alpha$ viscosity determines whether the energy due to nuclear reactions is comparable to that
due to the viscous effects \cite{mc00,zhang09} and Ohmic dissipation or not. From equation (\ref{modion}) we can write down 
the entropy per unit volume, namely the entropy density, of the flow as
\be
s=\int\frac{q_{\rm vis}+q_{\rm mag}+q_{\rm Nex}-q_{\rm rad}-q_{\rm Nend}}{\vartheta T}\,dx.
\label{s}
\ee
Now the turbulent kinematic viscosity 
$\nu={\vartheta_t l_t}/{3}$,
when $\vartheta_t$ and $l_t$ are respectively the turbulent Eddy velocity and the size of Eddy, which are respectively
to be scaled linearly with sound speed ($c_s$) of the disk and $h$ and hence
\be
\nu=\frac{\alpha_\vartheta c_s\,\alpha_l h}{3}=\alpha c_s h,
\label{eta}
\ee
where $\alpha_\vartheta$, $\alpha_l$ and hence $\alpha$ are constants. 
Therefore, the shear viscosity is given by 
\be
\eta=\alpha \rho c_s h.
\label{eta}
\ee

In an optically thin flow, often $q_{\rm rad}$ is approximately chosen to be proportional to $q_{\rm vis}+q_{\rm mag}$, in order to
understand the hydrodynamic properties of the flow without much loss of generality, such that
$q_{\rm rad}=(1-f)(q_{\rm vis}+q_{\rm mag})$ (see, e.g., \cite{ny94,mc00,mukhraj10-2}), when $f$ is a constant.
For an ADAF $f=1$ strictly (and $\dot{M}<<\dot{M}_{\rm Edd}$, when $\dot{M}_{\rm Edd}$ is the Eddington accretion rate) 
and for a GAAF $0<f\le 1$ (and $\dot{M}$ is larger than that in an ADAF \cite{mukhraj10}). Note that $f=0$ for an optically
thick Keplerian accretion disk. Hence, following previous work (e.g. \cite{mg03,mukhraj10,mukhraj10-2}),
we can write
\be
q_{\rm vis}+q_{\rm mag}-q_{\rm rad}=f\,(q_{\rm vis}+q_{\rm mag})
=f\left(\alpha(I_{n+1}c_s^2+I_n\vartheta^2)\rho\frac{d}{dx}\left(\frac{\lambda}{x}\right)+\frac{3|\vec{B}|^2\vartheta}{16\pi x}\right),
\label{qvis}
\ee
where $I_n = (2^n n!)^2 /(2n + 1)!$. Therefore, if $q_{\rm mag}, q_{\rm Nex}/f, q_{\rm Nend}/f<<q_{\rm vis}$, which we
name as viscous regime, $\eta/s$ does not explicitly depend on
$\alpha$ (and $\eta$) \footnote{In general, $\eta/s\sim T\eta/(W_{x\phi}^2/\eta)=
T/\frac{d}{dx}\left(\lambda/x\right)$, when $q_{\rm Nex}, q_{\rm Nend}<<fq_{\rm vis}$.}, 
depends only
on the hydrodynamic quantities whose numerical values, in a fixed range of radii, do not vary significantly 
in the optically thin regime with the change of $\alpha$. In the viscous regime, $\eta/s$ simply scales
with $M$. Note that viscosity depends on the size of the system, which in turn is determined 
by the mass of the black hole. Hence, only for a primordial black hole of $M\lsim 10^{16}$gm, $\eta/s$ is close to its theoretical
lower bound, as was suggested earlier \cite{sm11}. 
It was then also suggested that $\alpha$ has to be 
very small in an ADAF to obtain $\eta/s\sim\hbar/4\pi k_B$ for an astrophysical black hole. 
However, for a small $\alpha$, viscous dissipation (and then $s$) in an ADAF (or generally
in a GAAF) should be small as well, as is apparent from equations (\ref{modion}) and (\ref{qvis}). Hence, the ratio of $\eta$ 
to $s$ should not be invariably small, as described above, in the viscous regime. Therefore, while the earlier work \cite{sm11}
is very interesting and the first attempt to evaluate $\eta/s$ in an astrophysical flow, the flow
(and hence any conclusion)
should be amended with extra physics. In other words, the conclusion 
is incomplete.

Indeed $\eta/s$ could be small at a small $\alpha$ when $s$ does not decrease in 
that extent with the decrease of $\alpha$.
However, this is possible only when $q_{\rm mag}>>fq_{\rm vis}$ or/and $q_{\rm Nex}~{\rm or/and}~q_{\rm Nend}>>fq_{\rm vis}$, which we
name magnetic and nuclear regimes respectively. The magnetic regime essentially corresponds to the Blandford-Payne mechanism in the
hard spectral state,
when accretion takes place via magnetic stress, independent of the presence of shear/turbulent viscosity \cite{bland}.
This plausibly corresponds to the case of accretion flows with outflows/jets, e.g. Sgr~A$^*$, hard states
of GRS~1915+105.
Hence, even at a negligibly small $\alpha$ accretion is possible with a negligible $\eta/s$.
On the other hand, the nuclear reactions and 
corresponding energies depend on the density and temperature of the system. 
As shown earlier \cite{m03,mukhraj10,chak1,chak2,das1,das2} that even in the inviscid limit, 
temperature and density of the optically thin flows remain as high as those of a viscous accretion flow
for a fixed $M$, which successfully explain many observation \cite{chak3,chak4}.
Indeed, it was shown \cite{mc01,mc00} that significant nuclear energy ($\sim 10^{15}$ ergs/cc/sec)
throughout the disk may be released/absorbed in an inviscid flow which could lead to disk instability. Hence, from equation (\ref{s})
it is naturally possible that $s$ is large even at a small $\alpha$. 
Therefore, $\eta/s\sim \hbar/4\pi k_B$ also plausibly in the nuclear regime of the accretion flow, when 
$|q_{\rm Nex}-q_{\rm Nend}|\sim |q_{\rm Nex}|~{\rm or}~|q_{\rm Nend}|$ which could be the case in the hot 
accretion flows \cite{cm99,mc00,mc01}.

\section{Solutions and {\large  $\eta/s$}}\label{res}

\begin{figure} [ht]
\includegraphics[width=0.82\textwidth,angle=0]{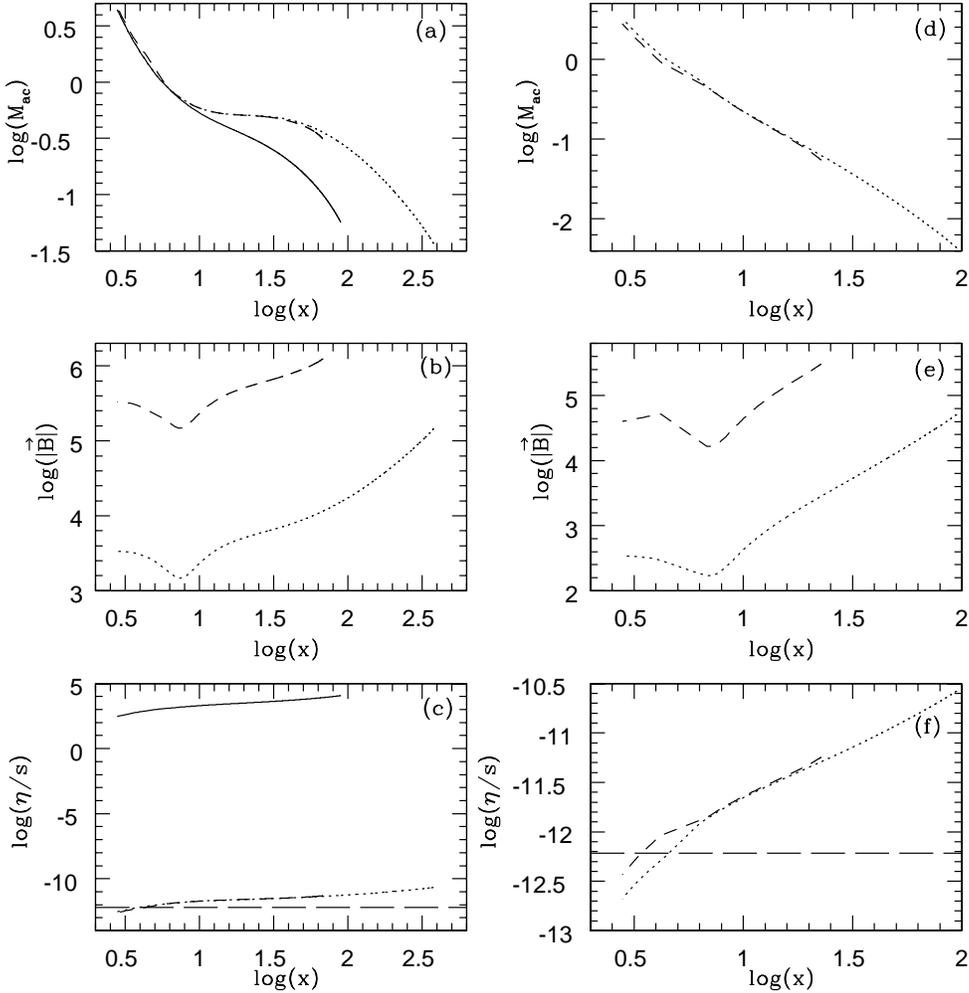} \\
\caption{Variations of (a) Mach number, (b) magnetic field in G, 
(c) $\eta/s$ in CGS unit when the horizontal line (long-dashed) refers to $\hbar/4\pi k_B$,
as functions of flow radius for $M=10M_\odot$ at $\dot{M}=\dot{M}_{\rm Edd}$.
Solid line corresponds to the flow in viscous regime and 
dotted and dashed lines correspond to the flows in magnetic regime.
(d), (e), (f) describe same as in (a), (b), (c) respectively, except
for $\dot{M}=0.01\dot{M}_{\rm Edd}$. See Table 1
for other details.}
\label{m10b}
\end{figure}

Let us understand the above physics quantitatively. Figures \ref{m10b}a,b,c show hydrodynamic/hydromagnetic properties 
of a disk around a $10M_\odot$ nonrotating black hole in the magnetic regime, accreting mass at the rate 
$\sim\dot{M}_{\rm Edd}$.
In this case of weakly viscous hot flow with $\alpha\sim5\times 10^{-17}$ (but not zero exactly), $\eta/s$ 
is shown to be close to the theoretical lower bound $\hbar/4\pi k_B$. 
For comparison, $\eta/s$  is also shown 
for a viscous regime, which is quite large for $\alpha=0.05$ and $f=0.1$. We also show the variations of
Mach number and magnetic field in the disk for the cases mentioned above for reference. Interestingly,
increase of magnetic field in the magnetic regime increases rate of transport which further renders
advancing the Keplerian disk towards the black hole, making the sub-Keplerian regime, which is shown here,
smaller. Figures \ref{m10b}d,e,f depict the properties for $\dot{M}\sim10^{-2}\dot{M}_{\rm Edd}$ in the 
magnetic regime. A very small (but nonzero) $\alpha$ ($\sim2\times 10^{-19}$) reveals $\eta/s$ to exhibit
the theoretical lower limit. The impact of change of the magnetic field strength to the disk is same
as that in the high accreting case discussed above.
Interestingly, independent of the accretion rate, magnetic field $\gsim$kG is able to transport matter efficiently 
in the sub-Keplerian flow in absence of viscosity, as was already argued by Blandford \& Payne \cite{bland}
in the Keplerian, self-similar framework. A smaller field strength would also transport matter as well,
except with the increase of size of the sub-Keplerian regime.

\begin{figure} [ht]
\includegraphics[width=0.82\textwidth,angle=0]{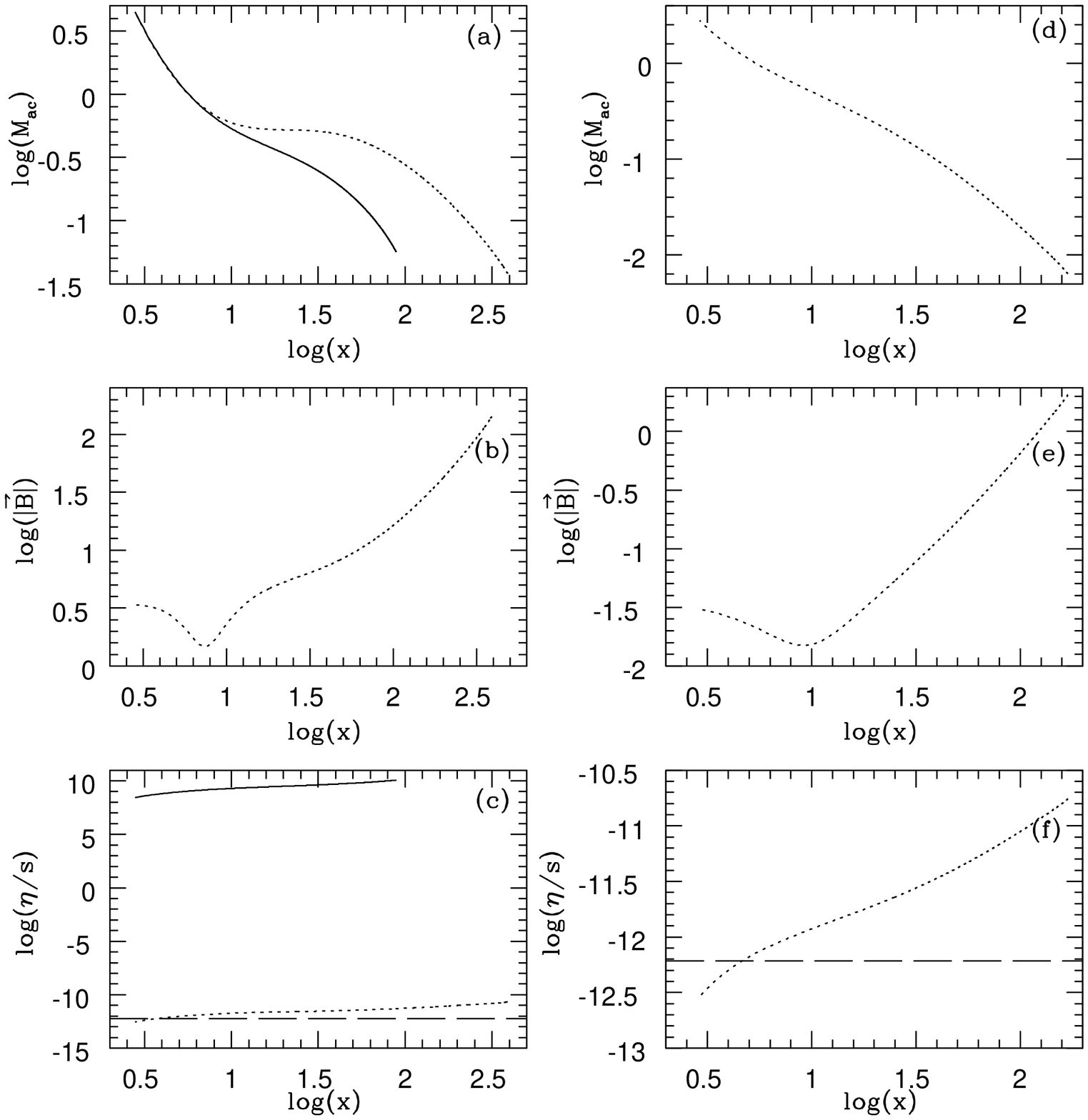} \\
\caption{Variations of (a) Mach number, (b) magnetic field in G, 
(c) $\eta/s$ in CGS unit when the horizontal line (long-dashed) refers to $\hbar/4\pi k_B$,
as functions of flow radius for $M=10^7M_\odot$ at $\dot{M}=\dot{M}_{\rm Edd}$.
Solid line corresponds to the flow in viscous regime and 
dotted line corresponds to the flow in magnetic regime.
(d), (e), (f) describe same as in (a), (b), (c) respectively, except
for $\dot{M}=10^{-4}\dot{M}_{\rm Edd}$. See Table 2
for other details.}
\label{me7b}
\end{figure}

Figure \ref{me7b} is devoted for hydrodynamic/hydromagnetic properties 
around a $10^7M_\odot$ nonrotating black hole in the magnetic regime. While Figs. \ref{me7b}a,b,c 
correspond to $\dot{M}\sim\dot{M}_{\rm Edd}$, Figs. \ref{me7b}d,e,f correspond to
$\dot{M}\sim10^{-4}\dot{M}_{\rm Edd}$. 
In either of the cases, very weakly viscous ($\alpha\sim5\times 10^{-23}$ for $\dot{M}\sim\dot{M}_{\rm Edd}$
and $\alpha\sim 10^{-25}$ for $\dot{M}\sim10^{-4}\dot{M}_{\rm Edd}$) hot flows reveal small $\eta/s$, 
close to its theoretical lower bound $\hbar/4\pi k_B$. 
For comparison, $\eta/s$  is also shown 
for a viscous, high accreting regime which is quite large for $\alpha=0.05$ and $f=0.5$. 
Interestingly, magnetic field $\gsim$mG is able to transport matter efficiently in the sub-Keplerian 
flow around a supermassive black hole in absence of viscosity, as was already argued by Blandford \& Payne \cite{bland}
in the Keplerian, self-similar framework. As is the case for a stellar mass black hole, 
a smaller field strength would also transport matter, except the fact that the Keplerian to sub-Keplerian transition
would arise at a larger radius.

\begin{figure} [ht]
\includegraphics[width=0.62\textwidth,angle=0]{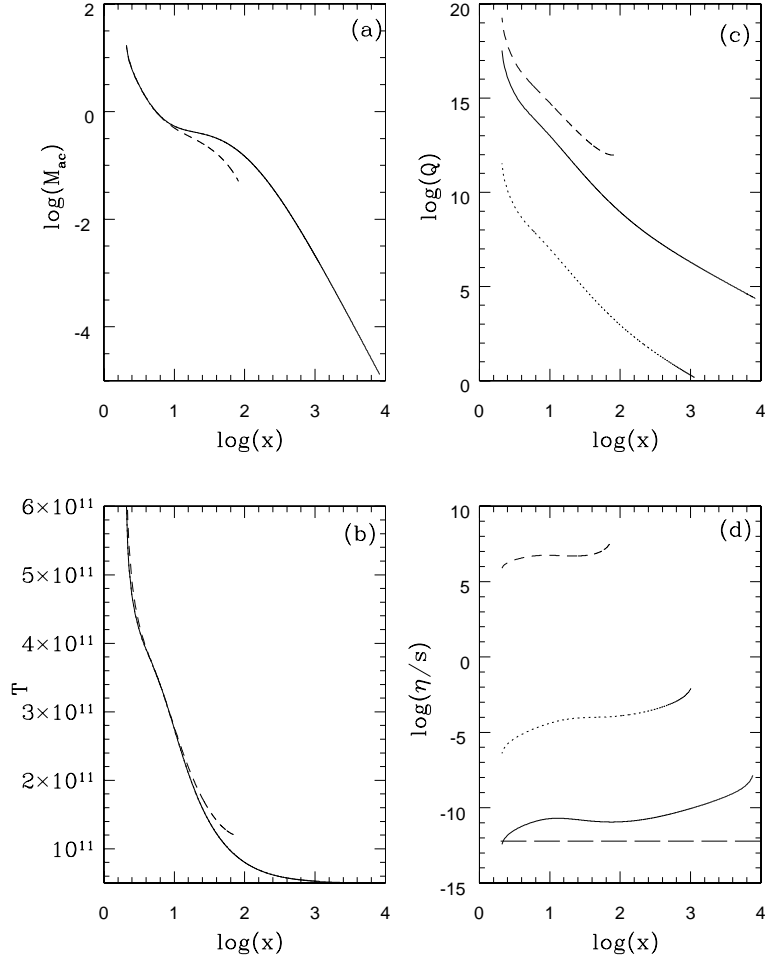} \\
\caption{Variations of (a) Mach number, (b) temperature in K, (c) energy density in CGS unit, 
(d) $\eta/s$ in CGS unit,
as functions of flow radius for $M=10M_\odot$ at high mass accretion rates. 
Solid and dotted lines correspond to the flows of very high 
and high nuclear regimes respectively and dashed line corresponds to the flow of viscous regime. See Table 3
for other details.}
\label{rad}
\end{figure}

Let us also look into the nuclear regime.
Figure \ref{rad} shows hydrodynamic properties of a disk around a $10M_\odot$
nonrotating black with
$\dot{M}=0.5\dot{M}_{\rm Edd}$ in the nuclear regime.
In the cases of weakly viscous hot flow, $\eta/s$ 
is shown to be inversely proportional to the 
nuclear energy released per unit volume. When the ratio of nuclear energy per unit volume 
to viscous energy per unit volume is chosen to be $Q_{\rm N}/Q_{\rm vis}\sim 10^{16}$
(which corresponds to $Q_{\rm max}\sim 3.4\times 10^{17}$ erg/sec, depicted by the solid line), $\eta/s$ appears
to be close to the theoretical lower bound $\hbar/4\pi k_B$. Note that the production of nuclear energy also
depends on the residence time of matter in the disk which is determined by $\alpha$, polytropic
constant $\gamma$, initial
mass fraction of the inflowing isotopes etc.
For comparison, $\eta/s$  is also shown 
for a viscous regime, which is quite large for $\alpha=0.05$ and $f=0.1$. We also show the variations of
Mach number and temperature in the disk for the cases mentioned above for reference. Interestingly,
the latter profiles overlap each other in the nuclear regime, independent of $Q_{\rm N}/Q_{\rm vis}$,
and are very similar to that in the viscous regime.
This also verifies that the Mach number and temperature do not depend much on $\alpha$, in 
the optically thin flows.

\begin{figure} [ht]
\includegraphics[width=0.62\textwidth,angle=0]{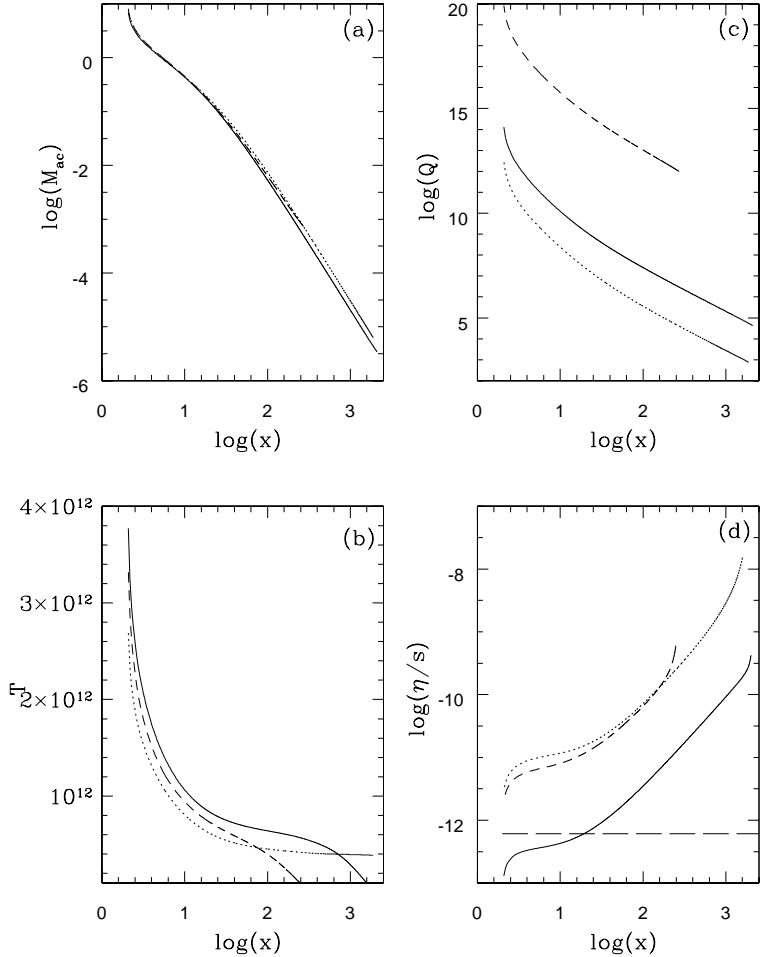} \\
\caption{Variations of (a) Mach number, (b) temperature in K, (c) energy density in CGS unit, 
(d) $\eta/s$ in CGS unit,
as functions of flow radius for $M=10M_\odot$. Solid and dotted lines correspond to the flows 
of very high and high nuclear regimes respectively at low mass accretion rates 
and dashed line corresponds to 
the flow of a nuclear regime at a high mass accretion rate. See Table 3
for other details.}
\label{mix}
\end{figure}

In Fig. \ref{mix}, we show a couple of sets of solutions for highly sub-Eddington hotter flows and a solution set
for a low angular momentum, high mass accreting, hot flow, in the nuclear regimes. Interestingly, a very 
low angular momentum ($\lambda\sim 1$), gas dominated, hot accretion flow with $\dot{M}=10^{-4}\dot{M}_{\rm Edd}$
(solid line) violates the theoretical lower limit of $\eta/s$. This restricts $\lambda$ of the flow. If $\lambda$ 
increases from $1$ to $2$, the flow becomes relatively cooler due to decrease of radial velocity 
which restricts nuclear energy, rendering
slightly larger $\eta/s>\hbar/4\pi k_B$ (dotted line). A flow with similar $\lambda$ ($\sim 1.6$) and $\alpha$, but
$\dot{M}=0.5\dot{M}_{\rm Edd}$ and hence revealing a radiation dominated accretion (dashed line) flow, exhibits however  
a similar $\eta/s > \hbar/4\pi k_B$. Note that while the nuclear energy among the three cases considered
here is highest in the last flow, due to presence of radiation the cooling efficiency therein is maximum,
which renders its $\eta/s$ above the lower limit throughout.


\vskip0.2cm {\centerline{\large Table 1}}{\begin{center} Parameters for the results shown
in Figure \ref{m10b} when $M=10M_\odot$ \end{center}}

\begin{center}
\begin{tabular}{|c|c|c|c|c|c|c|c|c|}

\hline
$\dot{M}$  & $\gamma$ & $\lambda$ & Regime & $|\vec{B}|$ in CGS 
& $f$ & Figure, line \\
in $\dot{M}_{\rm Edd}$ units &  & in inner disk   & &  in inner disk
&  &  \\
\hline

$1$ & $1.335$ & $3.2$ & Magnetic, $\alpha=5\times 10^{-17}$ & $1.85\times 10^3$G & $0.5$ & Figs. \ref{m10b}a,b,c, dotted\\
$1$ & $1.335$ & $3.2$ & Magnetic, $\alpha=5\times 10^{-17}$ & $1.85\times 10^5$G & $0.5$ & Figs. \ref{m10b}a,b,c, dashed\\
$1$ & $1.335$ & $3.2$  & Viscous, $\alpha=0.05$ & $-$ & $0.1$ & Figs. \ref{m10b}a,c, solid \\

\hline

$0.01$ & $1.5$ & $3.25$ & Magnetic, $\alpha= 2\times10^{-19}$ & $220$G & $0.5$ & Figs. \ref{m10b}d,e,f, dotted\\
$0.01$ & $1.5$ & $3.25$ & Magnetic, $\alpha= 2\times10^{-19}$ & $2.2\times 10^4$G & $0.5$ & Figs. \ref{m10b}d,e,f, dashed\\

\hline
\end{tabular}

\end{center}
\vskip0.2cm {\centerline{\large Table 2}}{\begin{center} Parameters for the results shown
in Figure \ref{me7b} when $M=10^7M_\odot$ \end{center}}

\begin{center}
\begin{tabular}{|c|c|c|c|c|c|c|c|c|}

\hline
$\dot{M}$  & $\gamma$ & $\lambda$ & Regime & $|\vec{B}|$ in CGS 
& $f$ & Figure, line \\
in $\dot{M}_{\rm Edd}$ units &  & in inner disk  & &  in inner disk
&  &  \\
\hline

$1$ & $1.335$ & $3.2$ & Magnetic, $\alpha=5\times 10^{-23}$ & $1.85$G & $0.5$ & Figs. \ref{me7b}a,b,c, dotted\\
$1$ & $1.335$ & $3.2$  & Viscous, $\alpha=0.05$ & $-$ & $0.5$ & Figs. \ref{me7b}a,c, solid \\

\hline

$10^{-4}$ & $1.53$ & $2.6$ & Magnetic, $\alpha= 10^{-25}$ & $0.0185$G & $0.5$ & Figs. \ref{me7b}d,e,f, dotted\\

\hline
\end{tabular}

\end{center}

\clearpage
{\centerline{\large Table 3}}{\begin{center} Parameters for the results shown
in Figures \ref{rad} and \ref{mix} when $M=10M_\odot$ \end{center}}

\begin{center}
\begin{tabular}{|c|c|c|c|c|c|c|c|c|}

\hline
$\dot{M}$ in $\dot{M}_{\rm Edd}$ units & $\gamma$ & $\lambda$ & Regime & $Q_{\rm max}$ in CGS 
& $f$ & Figure, line \\
\hline

$0.5$ & $1.34$ & $3.2$ & Nuclear & $3.4\times 10^{17}$ & $-$ & Fig. \ref{rad}, solid\\
$0.5$ & $1.34$ & $3.2$ & Nuclear & $3.4\times 10^{11}$ & $-$ & Fig. \ref{rad}, dotted \\
$1$ & $1.335$ & $3.2$ (inner) & Viscous, $\alpha=0.05$ & $1.8\times 10^{19}$ & $0.1$ & Fig. \ref{rad}, dashed \\

\hline
$0.0001$ & $1.53$ & $1$ & Nuclear & $1.3\times 10^{14}$ & $-$ & Fig. \ref{mix}, solid\\
$0.0001$ & $1.53$ & $2$ & Nuclear & $3\times 10^{12}$ & $-$ & Fig. \ref{mix}, dotted \\
$0.5$ & $1.34$ & $1.6$ & Nuclear & $8.4\times 10^{19}$ & $-$ & Fig. \ref{mix}, dashed \\

\hline
\end{tabular}

\end{center}

%
%
%
%
%

\section{Summary and discussion}\label{sum}

We have shown the astrophysical existence of small $\eta/s$, close to its theoretical lower bound.
As our physical inputs, governing the values of 
$\eta/s$, are in accordance with observed data, this could be argued as a natural evidence for 
$\eta/s$ close to its lower bound. This is possible in the cases of hot, optically thin,
accretion flows. More precisely, in order to have a small $\eta/s$, flow should be
magnetically dominated and/or producing huge energy/entropy other than that due to viscosity. 

The celebrated Blandford-Payne mechanism \cite{bland} was already shown to produce accretion
via magnetic stress (which helps in producing outflows/jets and removing angular momentum) 
in absence of viscosity in the framework of Keplerian, self-similar flow. 
We have employed this idea, but relaxing self-similar assumption, in order to obtain 
$\eta/s$ close to $\hbar/4\pi k_B$ in our accretion flows. We have found that in the 
presence of magnetic field of the order of mG to kG, the disk flows exhibit this value
of $\eta/s$ for supermassive to stellar mass black holes. 
Observed hard X-rays in Cyg~X-1, some temporal
classes (e.g. $\chi$) of GRS~1915+105, Sgr~A$^*$ are observed to be associated with jets which 
argues for the sources to exhibit strong magnetic fields. Hence, those sources could be 
considered to be natural sites for $\eta/s\sim \hbar/4\pi k_B$.

Additionally, the flow exhibiting nuclear 
energy could be a natural site for small $\eta/s$. Nuclear
energy depends on the density and temperature of the flow. 
The temperature in an optically thin, sub-Keplerian accretion flow, producing
observed hard X-rays, could be $\gsim 10^9$K.
Therefore, eventhough the density therein is small, very high
temperature would suffice for nuclear burning. We have recalled the previous work
discussing possible nuclear burning in accretion flows and its observational
evidences/consequences \cite{mc00,mc01,filho03,tao08,zhang09,tao11}. 
Based on them, we have modeled the hot, optically thin accretion flows for the present purpose and
shown how the nuclear energy in the flow determines the value of $\eta/s$. Thus
we argue that the flows producing hard X-rays in Cyg~X-1, some temporal
classes (e.g. $\chi$) of GRS~1915+105 could be hot enough to have $\eta/s\sim \hbar/4\pi k_B$,
atleast close to the black hole for certain nuclear regimes, when flows are necessarily sub-Keplerian due to the 
dominance of gravitational power of the black hole.
It will be now interesting to explore $\eta/s$ for other astrophysical flows, having,
e.g., strong magnetic field or/and high temperature and/or density. 

Finally, we end by enlightening maximum Reynolds number ($R_e$) to the flows, restricted
by the lower bound of $\eta/s$. $R_e$, which is the ratio of typical (characteristic) velocity times
length scale to the kinematic viscosity of the flow, can be recast as
$R_e=M_{ac}\,(x/h)/\alpha$. Now from Figs. \ref{m10b} and \ref{rad}, $M_{ac}$ close
to the black hole is $\sim 5-10$. However, the magnetic and nuclear regimes correspond to very low
but nonzero viscosity revealing $\alpha\gsim 10^{-16}$ for $\eta/s$ lower bound to 
be satisfied. As typically 
$h/x\sim 0.1-0.5$ for an optically thin, geometrically thick flow, for stellar
mass black holes $R_e\lsim 10^{17}$. If we look at Fig. \ref{me7b} showing 
$M_{ac}$ to be of the similar order in magnitude around supermassive black holes, however for smaller 
values of $\alpha$ ($\gsim 10^{-22}$) in larger accretion disks (as size scales with the mass of the 
black hole), we obtain $R_e\lsim 10^{23}$ for those flows. 

{\bf Acknowledgments} : The author would like to thank Zoltan Fodor for comments and
suggestions along with bringing in his attention to important references. Thanks are also due 
to Upasana Das and Aninda Sinha for discussion.
This work was partly supported by an ISRO grant ISRO/RES/2/367/10-11.

\end{document}